\documentclass[12pt]{iopart}
\usepackage{graphicx}

\begin{document}
\title[]{Crossing $w=-1$ by a single  scalar field coupling with matter
and the observational constraints}

\author{M L  Tong$^{1,2}$, Y  Zhang$^{1}$ and Z W  Fu$^{1}$}

\address{$^1$Key Laboratory for Researches in  Galaxies  and Cosmology, Chinese Academy of Sciences,
Department of  Astronomy,   University of Science and Technology of
China, Hefei,  230026, China}
\address{$^2$Korea Astronomy and Space Science Institute, Daejon 305-348, Korea}
\ead{mltong@mail.ustc.edu.cn }

\begin{abstract}
Motivated by Yang-Mills dark energy model, we propose
a new model by introducing a logarithmic correction.
we find that this model can avoid the coincidence problem
naturally and gives an equation of state $w$ smoothly crossing $-1$
if an interaction between dark energy and dark matter exists.
It  has a stable tracker solution as well.
To confront with observations
based on the combined data of SNIa, BAO, CMB and Hubble parameter,
we obtain the best fit values of the parameters with $1\sigma, 2\sigma,
3\sigma$ errors for the  noncoupled model:
$\Omega_m=0.276\pm0.008^{+0.016+0.024}_{-0.015-0.022}$, $h=0.699\pm0.003\pm0.006\pm0.008$,  and for
 the coupled model with a decaying rate $\gamma=0.2$:
$\Omega_m=0.291\pm0.004^{+0.008+0.012}_{-0.007-0.011}$,
$h=0.701\pm0.002\pm0.005\pm0.007$. In particular, it is found  that
the non-coupled model has a dynamic evolution almost
undistinguishable to $\Lambda$CDM at the late-time Universe.

\

\end{abstract}

\maketitle

\section{ Introduction}

The accelerating expansion has been supported by observations of
supernova observations  \cite{Riess1,Perlmutter},
 cosmic microwave background radiation (CMB),
and the large scale structure through the baryon acoustic
oscillation (BAO). However, the physical origin of acceleration is
still a challenging mystery. Within the framework of general
relativity, this can be interpreted by a cosmic dark energy with
negative pressure. The simplest dark energy model is the
cosmological constant ($\Lambda$CDM), where the vacuum energy is
responsible for the accelerating expansion. Whereas, it suffers from
two problems degenerately. The first is the fine-tuning problem: The
observed vacuum energy density of order $\sim10^{-47}$ GeV$^4$ is
about $10^{121}$ orders of magnitude smaller than the value expected
by quantum field theory for a cut-off scale being the Plank scale,
and is still about $10^{44}$ orders smaller even for a cut-off scale
being the QCD scale \cite{Copeland}. The second is the coincidence
problem: The conditions in the early Universe have to be set very
carefully in order for the energy density of the vacuum and that of
the matter to be comparable today. To solve these problems,
abundances of dynamical dark energy models have been proposed, such
as scalar field
\cite{Ratra1,Zlatev,Steinhardt,Ferreira,Dodelson,Carvalho}, vector
field  \cite{Zhang94,Zhang1,Kiselev,Picon}, and decaying vacuum
energy \cite{Borges,Tong2}, Holographic dark energy model
\cite{Holographic,Holographic2,Holographic3,Holographic4}
 and so on (see \cite{Copeland} for details).
In  quintessence of scalar field, the Lagrangian density has a
standard form $\mathcal{L}=\frac{1}{2}(\nabla \phi)^2 -V(\phi)$,
which contains a canonical kinetic term and a potential term. Some
particular  forms of the potentials  of quintessence \cite{Ratra1,Zlatev,Steinhardt}
or the coupled quintessence with dark matter \cite{Amendola,Chimento,Gonzalez},
 can not only
lead to the late time acceleration of the Universe, but also avoid
the coincidence problem.

 As well known, the quintessence always has an equation of state (EoS)
$w$ larger than $-1$. However, there have been some preliminary
evidences that the current value of $w$  may be less than $-1$, as
indicated from observations
\cite{Corasaniti,Alam,Astier,Conley,Vasey,Davis,Freedman}. On the other hand,
the
phantom field with a negative kinetic energy \cite{Caldwell}
 predicts a  EoS  being always smaller than $-1$.
 Furthermore,
another scalar field model with a noncanonical kinetic term called
K-essence was proposed \cite{Armen,Armen2,Chiba}. The Lagrangian is
generally taken to be $\mathcal{L}=f(\phi)g(X)$, where
$X=\frac{1}{2}(\nabla \phi)^2$. In \cite{Chiba}, $f(\phi) \propto
\phi^{-\alpha}$ and $g(X)$ is a polynomial of $X$, leading to  $w
>-1$ for $\alpha>0$ as a quintessence and $w <-1$ for $\alpha<0$ as
a phantom. For the considerations of the EoS of dark energy tracking
that of the background, one  expect that the EoS of dark energy  may
be lager than $-1$ in early times and smaller than $-1$ in late
times, i.e., $w$ crosses $-1$. If this is supported by further
observations, many dark energy models would undergo a problem. Note
that, even in K-essence model,
  $w$ can not cross $-1$ for a fixed $\alpha$.
 Is it possible to have a transit from quintessence to phantom?
 The particular  interacting phantom dark energy could  give
 smooth transit from $w>-1$ to $w<-1$ \cite{Curbelo}.
 The non-minimal interaction between dark matter and dark
energy with a single scalar field could also make possible to do the crossing
of the phantom divide \cite{Gonzalez2}. Besides, based on
employment of two scalar fields, a class of models have been
proposed, thereby using extra degrees of freedom  \cite{Feng,Hu2,
zhaozhang}.

In this paper,  we propose to study a new dark energy model
described by a single scalar field, whose
Lagrangian density contains a logarithmic factor. We call it
{\it effective scalar field} (ESF) dark energy model.
 Since the kinetic energy
term is noncanonical,  it should  belong to a subclass of K-essence
models. This is inspired by our previous work on the quantum
effective Yang-Mills condensate (YMC) dark energy model
\cite{Zhang1,WZhao1,Zhang3,Xia,MLTong,SWang} with
$\mathcal{L}_{eff}\propto F\ln F$ for 1-loop case
\cite{Zhang94,Zhang1,WZhao1,Zhang3}, where $F$ is the squared gauge
field strength \cite{Pagels,Adler,Zhang94}. The nonlinear kinetic
terms appear generically in the effective action in string and
supergravity theories \cite{Gross}. The appearance of a logarithmic
correction in the field is generic for effective quantum theories,
e.g., the Coleman-Weinberg potential \cite{Coleman}, the effective
gravity \cite{Parker,Parker2}, as well as the effective Yang-Mills
field \cite{Pagels}. Since the nature of dark energy is still
unknown, in the following, we will investigate the phenomenological
properties of ESF model. Based on the observation of nearby galaxies
\cite{peebles}, an interaction between dark energy and dark matter
is favored since it would give a more rapid structure formation than
predicted by the  $\Lambda$CDM model \cite{Duran}. So, we will also
generally consider that an interaction  between dark energy and
matter exists. As will be seen, with one scalar field, the model
provides a smooth dynamical transit from quintessence to phantom,
with its $w $ going from $>-1$ at high reshifts to $<-1$ at low
redshifts, if the field decays into matter. All the physical
quantities involved in the model are smooth during the whole
dynamical evolution. Moreover,  the coincidence problem is also
avoided in this model, since it has a stable attractor solution.
But, unfortunately, the fine-tuning problem is still exist since a
model parameter has to be tuned to accord with the low density of
the dark energy component obtained through observations. We will
demonstrate these afore-mentioned points, and also carry out a joint
$\chi^2$ analysis for the  model, confronting it with recent
observations from SN Ia \cite{Kowalski,Hicken,Amanullah}, BAO peak
measurement of large scale structure from the Sloan Digital Sky
Survey (SDSS) \cite{Eisenstein} and the Two Degree Field Galaxy
Redshift Survey (2dFGRS) \cite{Percival},
 the shift parameter of CMB  \cite{Komatsu},
and the history of the Hubble parameter \cite{Stern,Riess2,Gazta}.
Throughout this paper, we adopt a unit with $c=1$.
 Greek indices $\mu, \nu,...$ range over
0 to 3, and Latin indices $i, j,...$ range over 1 to 3.

\section{The effective scalar field model}

We consider  a spatially flat Universe described by  Friedmann-Robertson-Walker metric
 \begin{equation}
 ds^2=dt^2-a^2(t)\delta_{ij}dx^idx^j,
 \end{equation}
 where the scalar factor $a(t)$ is determined by the Friedmann equation:
 \begin{equation}
 H^2=\frac{8\pi G}{3}(\rho_{\phi}+\rho_m+\rho_r),
  \end{equation}
 where
 $\rho_\phi$, $\rho_m$ and $\rho_r$ represent
 energy density of dark energy, matter and radiation, respectively,
 and $H=\dot{a}/a$ is the Hubble parameter.
The dark energy is described by a
scalar field $\phi$  with a Lagrangian density
\begin{equation}
\label{Langrangian} {\cal L}_\phi
=\left(\frac{1}{2}(\nabla \phi)^2
     -V(\phi)\right)\ln\left|\frac{\frac{1}{2}
            (\nabla \phi)^2
     -V(\phi)}{\alpha e}\right|,
\end{equation}
where $(\nabla\phi)^2=g^{\mu\nu}\partial_\nu\phi\partial_\nu\phi$,
 $V(\phi)$ is a function  of  $\phi$, $\alpha$
is a scale of energy density to be fixed by observations, and $\ln
e=1$. ${\cal L}_\phi$ in Eq.(\ref{Langrangian}) is formally similar
to the Lagrangian density of the 1-loop effective YMC dark energy
model \cite{Zhang1}. Assuming $\phi$ is homogeneous and
isotropic, and only depends on time, i.e., $\phi=\phi(t)$. The action is given by
\begin{equation} \label{action}
S=\int d^4x\sqrt{-g}{\cal L}_\phi,
\end{equation}
where $g\equiv {\rm det}(g_{\mu\nu})$. The variation of the action (\ref{action})
with respect to $\phi$ gives
 \begin{equation}\label{fieldeq}
\ddot{\phi}+\left( 3\frac{\dot{a}}{a}
+\frac{\dot{\varepsilon}}{\varepsilon}\right)\dot{\phi}
                +\frac{dV}{d\phi}=0,
\end{equation}
where a dot denotes $d/dt$, and
$\varepsilon\equiv\ln\left|(\frac{1}{2}\dot{\phi}^2-V)/\alpha\right|$.
Note that, Eq.(\ref{fieldeq}) differs from that of the quintessence model
by the extra term $(\dot{\varepsilon}/\varepsilon)\dot\phi$ .
When a coupling exists between ESF and  matter,
their dynamical evolution equations  are given by
\begin{eqnarray}\label{phicoupling}
&&\dot{\rho}_\phi+3H(\rho_\phi+p_\phi)=-\Gamma
\rho_\phi,\\
\label{mcoupling}
&&\dot{\rho}_m+3H \rho_m =\Gamma \rho_\phi,
\end{eqnarray}
where
 $\Gamma$ denotes the energy transformation rate from ESF to matter.
 For simplicity, we assume $\Gamma$ is a constant.
The radiation is an independent component
and evolves as $\rho_r(t)\propto a^{-4}$.

In the following we focus on
the simple case of $V=0$.
The energy density and pressure are easily gained by the variation
of action (\ref{action}) with respect to $g^{\mu\nu}$:
  \begin{equation}\label{rhophi1}
\rho_\phi= \alpha (\varepsilon+1)e^\varepsilon,
        ~~~~~ p_\phi=\alpha (\varepsilon-1)e^\varepsilon,
\end{equation}
with
$\varepsilon=\ln{(\frac{1}{2}\dot{\phi}^2}/\alpha)>-1$
required by  $\rho_\phi>0$.
It is easy to prove that, for the non-coupled case ($\Gamma=0$),
Eq. (\ref{phicoupling}) reduces to Eq. (\ref{fieldeq}) with the help
of Eq. (\ref{rhophi1}).
The EoS is given by
\begin{equation}\label{eos}
w=\frac{p_\phi}{\rho_\phi}=\frac{\varepsilon-1}{\varepsilon+1}.
\end{equation}
In high energy limit with  $\varepsilon\gg1$, $w\rightarrow1$,
different from the high energy behavior $w\rightarrow {1}/{3}$
of the YMC model \cite{Zhang1}.
At  the critical point $\varepsilon=0$, one has $w=-1$.
Furthermore, $w<-1$ will be arrived when $-1<\varepsilon<0$.
Introducing dimensionless  $x\equiv \rho_m/\alpha$ and $r\equiv
\rho_r/\alpha$, Eqs. (\ref{phicoupling}) and (\ref{mcoupling}) read
as
 \begin{eqnarray}  \label{scalar}
&&\varepsilon'+\frac{6\varepsilon}{\varepsilon+2}
          +\frac{\gamma(\varepsilon+1)}{\varrho(\varepsilon+2)}=0,\\
   \label{matter}
&&x'+3x-\frac{\gamma}{\varrho}(1+\varepsilon)e^{\varepsilon}=0,
\end{eqnarray}
where
$'\equiv {d}/{dN}$
with $N\equiv\ln a(t)$,
$\gamma\equiv\Gamma/
         \left(\frac{8\pi G \alpha}{3}\right)^{\frac{1}{2}}$
is the dimensionless decaying rate,
and
 $\varrho \equiv [x+r+e^{\varepsilon}(\varepsilon+1)]^{1/2}$.
Given initial values ($\varepsilon_i$, $ x_i$),
Eqs. (\ref{scalar}) and (\ref{matter}) can be solved
 for each $\gamma$. $\gamma>0$ means that dark energy decays
 into matter, and vice versa.
In this paper, we assume that $\gamma$ is positive,
since $\gamma<0$ will lead to
a negative matter density in the future, which
 is unacceptable from the point of view of physics.

\begin{figure}
\centerline{\includegraphics[width=12cm]{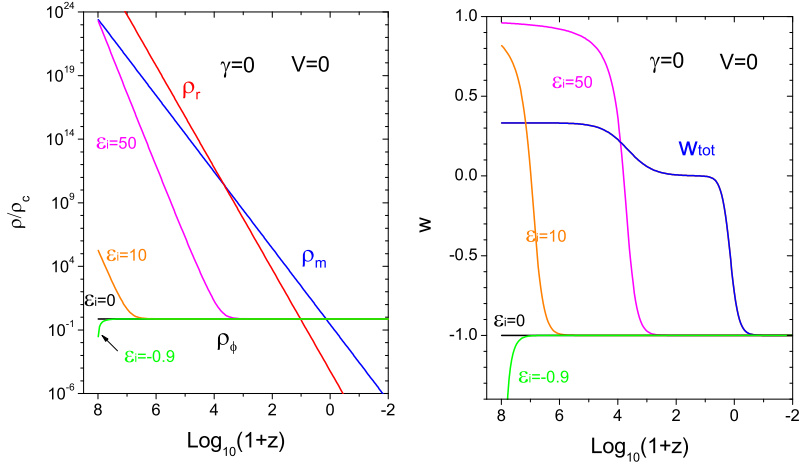}}
 \caption{\label{fig1}
 (a)  $\rho_\phi(z)$, $\rho_m(z)$, and $\rho_r(z)$
 in the non-coupled ESF model.
 For various initial values
$\varepsilon_i$ the current status  $\Omega_\phi=0.73$ and
$\Omega_m=0.27$ is always achieved.
(b) $w(z)$ does not cross $-1$ without the coupling.}
\end{figure}

Firstly, we  would like to discuss the simple case of
the non-coupled $(\gamma=0)$ ESF model. To ensure  the
standard  cosmology not being spoiled by the presence of dark
energy, one may take $\rho_{\phi i}/ \rho_{r i} \le 10^{-2}$ at
$z_i\simeq10^8$. The outcome is that, for any $\varepsilon_i$ in a
wide range $-1 < \varepsilon_i \le 55$ corresponding to $\rho_{\phi
i}$ ranging over almost infinity orders of magnitude, the current status
($\Omega_\phi, \Omega_m ) \simeq (0.73, 0.27$) is always attained.
So the  coincidence problem is solved at the price of choosing a
fixed $\alpha$.
As can be seen
 in Fig.\ref{fig1} (a), different initial values of $\varepsilon_i=-0.9,0,10,50$,
 lead to the same density of the ESF at present time.
 During earlier stages the decreasing $\rho_\phi(z)$
is subdominant to $\rho_r(z)$ and $\rho_m(z)$.
Note that $\rho_\phi(z)$ always levels off at a certain time,
earlier for a smaller $\varepsilon_i$.
Then, it surpasses $\rho_r$ at $z\sim 10$ and
surpasses $\rho_m$ at $z\sim 1$, respectively.
  Whereas, the matter density  $\rho_m(t)$
 evolves independently  as $\rho_m(t)\propto a(t)^{-3}$,
 since it does not couple with the ESF.
Fig.\ref{fig1} (b) shows that   $w$  will decrease and increase with time and approach  $-1$ at the late-time for $\varepsilon_i>0$ and $\varepsilon_i<0$, respectively. Moreover, for
$\varepsilon_i=0$, the ESF acts as the  $\Lambda$CDM model exactly. Note that, in the above three cases,
all the corresponding  $w$ will stay at $-1$ in the future and never cross $-1$.
  It can be understood as follows. For $\gamma=0$, Eq. (\ref{scalar}) has a
 solution:
 \begin{equation}\label{epsilon}
 \varepsilon^2 e^{\varepsilon}=A a^{-6},
 \end{equation}
 where $A$ is an integration constant determined by the initial conditions.
 Except the particular case that $\varepsilon=0$, $A$ is always positive, i.e.,
 the RHS of Eq. (\ref{epsilon}) will always be larger than 0. This indicates that
 $\varepsilon$ will never cross 0 no matter the initial value of $\varepsilon$
 is positive or negative. According to  Eq. (\ref{eos}), we know that $w$ can not cross
 $-1$ resulting from  $\varepsilon$ failing to cross $0$. In the special case of
 $\varepsilon_i=0$, $A$ is fixed to be zero. Thus, $\varepsilon=0$ will be kept all
 the time, i.e., $w=-1$ is constant. Since $\varepsilon^2e^{\varepsilon}$ decays with
 the expansion Universe as $\propto a^{-6}$, $\varepsilon\rightarrow0$ at $z=0$.
 Therefore,  $\rho_\phi(z=0)\simeq \alpha\simeq0.73\rho_c$, where $\rho_c$ is the critical density
 of the Universe.
 We find that, for the whole range of $\varepsilon_i$,
  the resulting dynamical evolution in the recent past $(z<30)$
is almost identical to that in $\Lambda$CDM
with deviations
$(\rho_\phi-\rho_\Lambda)/\rho_\Lambda < 10^{-4}$.
The total EoS is $w_{tot}=\sum\Omega_jw_j$, where $j$ stands for the ESF, matter and radiation, respectively. In the future ($N \rightarrow \infty$), one has $\Omega_r\rightarrow0$, $\Omega_m\rightarrow0$ and $\Omega_\phi\rightarrow1$. Hence, $w_{tot}=w=-1$, as shown in Fig.\ref{fig1} (b). That is, the Universe will do an exact de Sitter expansion, and there is no big rip event
 which some dark energy models would encounter. Note that, the interacting phantom
dark energy could also avoid the big rip event \cite{Curbelo}.

\begin{figure}
 \centerline{\includegraphics[width=12cm]{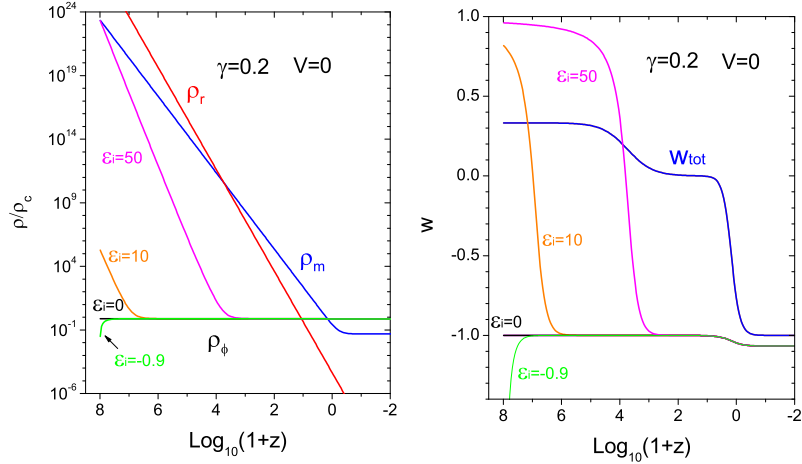}}
 \caption{\label{fig2}
 (a)  $\rho_\phi(z)$, $\rho_m(z)$, and $\rho_r(z)$
 in the coupled ESF model with $\gamma=0.2$.
 For various initial values
$\varepsilon_i$ the current status  $\Omega_\phi=0.73$ and
$\Omega_m=0.27$ is always achieved.
(b) $w(z)$ in the coupled model.
  Due to coupling, $w(z)$ is able to cross $-1$
  at low redshifts.}
\end{figure}

Secondly, as an explicit example of ESF, we  discuss
the coupled case with $\gamma=0.2$. The initial conditions for
the ESF and radiation are chosen  the same as the case of non-coupled ESF.
The initial condition for the matter is chosen a little differently from
the non-coupled case in order to ensure the  current status
($\Omega_\phi, \Omega_m ) \simeq (0.73, 0.27$).
As illustrated in Fig.\ref{fig2} (a), the coincidence problem is also
avoided  in this case.
The dynamic evolutions of the density of  the coupled  ESF are quite similar to those
in the non-coupled ESF. However,  $\rho_m$   levels  off around $z\sim 0$
and will   approach  a constant instead of decaying as $\propto a(t)^{-3}$.
This is caused by the coupling $\gamma\ne 0$.
Fig. \ref{fig2}(b) plots evolutions of the corresponding $w(z)$.  Due to the coupling, $w$
crosses $-1$, arrives at $w_0\simeq -1.05$ at present, and settles
down to a constant value $w\simeq-1.07$ in future. The influence of
the coupling $\gamma$ has been investigated, and computations show
that a greater $\gamma$  yields a larger matter fraction $\Omega_m$
and a smaller EoS $w_0$ at present.
We have also found an interesting
relation: $\Omega_\phi=-1/w$ as $N \rightarrow \infty$, which is
similar to YMC model \cite{zhao}. This implies that the total EoS
satisfies $w_{tot}=\Omega_\phi
w=-1$ as $N \rightarrow \infty$. Thus,  the coupled ESF model also predicts an
exact de Sitter expansion in future, and the big rip  event is avoided naturally.
The parameter $\alpha$ in this case
 are determined
by  setting $\rho_{\phi}\simeq0.73\rho_c$ at $z=0$,
leading  to $\alpha\simeq 0.76 \rho_c$.
Unfortunately, the particular choices of $\alpha$ being  the same order of magnitude
as $\rho_c$ let the ESF model still suffer from   the fine-tuning problem.

 We have  carried out an
analysis of dynamic stability of the set of Eqs. (\ref{scalar}) and
(\ref{matter}), and found that it has the fixed point $(\varepsilon_c,x_c)=(0,0)$
for $\gamma=0$ and
$(\varepsilon_c,x_c)=(-0.0323,0.0625)$ for $\gamma=0.2$, respectively, as $N\rightarrow \infty$.
Moreover,  any perturbations $\delta\varepsilon$ and $\delta x$ are
both   decay as $e^{-3N}$ for $\gamma=0$, and  decay as linear combinations of $e^{-3.3913N}$ and $
e^{-2.8372 N}$ for $\gamma=0.2$, respectively. Thus, the fixed points are stable, and then the attractor solutions of $\rho_\phi(t)$ are obtained   for the above two cases.
 Aside
the trajectory smoothness in phase space, the stability is
problematic in single scalar field models \cite{Vikman}.
 Moreover, all the physical
quantities in our models, such as $a(t)$, $\rho_\phi(t)$, and $w(t)$,
are smooth from the initial moment up to the future.

\section{Constraints from SN Ia, BAO,  CMB and Hubble data }

For a model of dark energy to be viable,
it needs to confront or be constrained with observational data.
Here we constrain the  $V=0$ model
with the latest observational data of the 557
SN Ia assembled in the Union2 compilation \cite{Amanullah},
the BAO measurement from SDSS \cite{Eisenstein} as well as 2dFGRS \cite{Percival},
the shift parameter of CMB from WMAP7 \cite{Komatsu}, and the
 the history of the Hubble parameter \cite{Stern,Riess2,Gazta}.

First, we compare the theoretical distance modulus to the observed ones
compiled in \cite{Amanullah}. The theoretical distance modulus is defined as
\begin{equation}
\mu_{th}(z)\equiv5\log_{10}D_L(z)+\mu_0,
\end{equation}
where $D_L(z)=H_0(1+z)\int_0^z \frac{dz'}{H(z')}$ is the  Hubble-free
luminosity distance in a spatially flat Universe, and
$\mu_0\equiv42.38-5\log_{10}h$ with $h$ the Hubble constant in
the unit of $100km/sec/Mpc$.
The late-time Hubble rate of the effective scalar model is given by
\begin{equation}
H(z)=H_0[\Omega_\phi(z)+\Omega_m(z)]^{1/2},
\end{equation}
where $\Omega_i(z)=\rho_i(z)/\rho_c$ for $i=\phi,m$.
Since the evolution for $z\leq10^3$ in this model is insensitive to
the initial conditions,
we choose $\varepsilon_i=10$ in the following calculations for concreteness.
For the  SN Ia data,
the  $\chi^2$ function is
\begin{equation}\label{chi2SN}
\chi^2_{SN}(p_s;\mu_0)
 =\sum_{i=1}^{557}\frac{[\mu_{th}(z_i)-\mu_{obs}(z_i)]^2}{\sigma_i^2},
\end{equation}
where $p_s$ stands for a set of parameters,
such as $\Omega_m$.
The nuisance parameter $\mu_0$ can be analytically marginalized
over \cite{Nesseris},
so that one actually minimizes $\chi^2_{SN}(p_s)$
instead of  $\chi^2_{SN}(p_s;\mu_0)$.

Next, the BAO is revealed by a distinct peak in the large scale
correlation function measured from the luminous red galaxies sample
of the SDSS at $z=0.35$ \cite{Eisenstein}, as well as in the 2dFGRS
at $z=0.2$ \cite{Percival}. The peaks can be associated to expanding
spherical waves of baryonic perturbations. Each peak introduces a
characteristic distance scale \cite{Eisenstein,Nesseris2} \begin{equation}
D_v(z_{BAO})=\left[\frac{z_{BAO}}{H(z_{BAO})}\left(\int_0^{z_{BAO}}
\frac{dz}{H(z)}\right)^2\right]^{1/3}. \end{equation} The observational date
from SDSS and 2dFGRS measurements yield
$D_v(0.35)/D_v(0.2)=1.736\pm0.065$ \cite{Percival}. The best fit
values for the  model are given by minimizing \cite{Duran,Xu} \begin{equation}
\chi_{BAO}^2(p_s)=\frac{([D_v(0.35)/D_v(0.2)]_{th}-
[D_v(0.35)/D_v(0.2)]_{obs})^2}{\sigma^2_{D_v(0.35)/D_v(0.2)}}, \end{equation}
where $\sigma^2_{D_v(0.35)/D_v(0.2)}=0.065$.

As discussed in \cite{dv,dv2},
the first peak of the CMB spectrum of anisotropies, $l_1$, is more suitable to be
used to test the interacting dark energy model    than the CMB shift parameter, $R\equiv \sqrt{\Omega_m}D_L(z_{rec})/(1+z_{rec})$ \cite{Bond,WangYu}, where $z_{red}=1091$ \cite{Komatsu}
is redshift of recombination.
Then, we  use $l_1$,  which
 is related to the angular scale, $l_A$, by \cite{Hu}
\begin{equation}
l_1=l_A(1-\delta_1),
\end{equation}
where
\begin{equation}
\delta_1=0.267\left(\frac{\bar\rho}{0.3}\right)^{0.1}
\end{equation}
with $\bar\rho\equiv \rho_r(z_{rec})/\rho_m(z_{rec})$  the density ratio of radiation
and matter at the time of recombination. The acoustic scale is defined as
\begin{equation}
l_A=\pi\int_0^{z_{rec}}\frac{dz}{H(z)}/\int_{z_{rec}}^{\infty}\frac{c_sdz}{H(z)},
\end{equation}
where the sound velocity is $c_s=(3+\frac{9\Omega_b a}{4\Omega_\gamma})$,
with $\Omega_b$ and $\Omega_\gamma$ the present density parameters of baryons
and photons, respectively.
With the observed position of the first peak $l_{1obs}=220.8\pm0.7$ \cite{Hinshaw},
 the $\chi^2$ for CMB is
\begin{equation}
\chi_{CMB}^2(p_s)=\frac{(l_{1th}-l_{1obs})^2}{\sigma_l^2},
\end{equation}
where $\sigma_l=0.7$.

Finally, the Hubble parameter as a function of redshift $z$ can be written as
\begin{equation}
H(z)=-\frac{1}{1+z}\frac{dz}{dt}.
\end{equation}
Then, once $dz/dt$ is known, $H(z)$ is obtained directly. Simon {\it et al.} \cite{Simon} and Stern
{\it et al.} \cite{Stern} obtained $H(z)$ in the range of $0\leq z \leq1.8$, using the differential
ages of passively-evolving galaxies and archival data. Recently, some high precision
measurements constrained $H(z)$ at $z=0$ from the observation of 240 Cepheid variables
of rather similar periods and metallicities \cite{Riess2}. Besides,  $H(z)$ at
$z=0.24$, $0.34$ and $0.43$ is obtained \cite{Gazta} by using the BAO peak position as
a standard ruler in the radial direction. We employ the twelve data in \cite{Riess2,Stern}
and the three data in \cite{Gazta}.
The best fit values of the model parameters from observational Hubble data are determined
by minimizing
\begin{equation}
\chi^2_{Hub}(p_s)=\sum_{i=1}^{15}\frac{[H_{th}(z_i)-H_{obs}(z_i)]^2}{\sigma^2(z_i)}
\end{equation}

\begin{figure}[t!]
\centerline{\includegraphics[width=6cm]{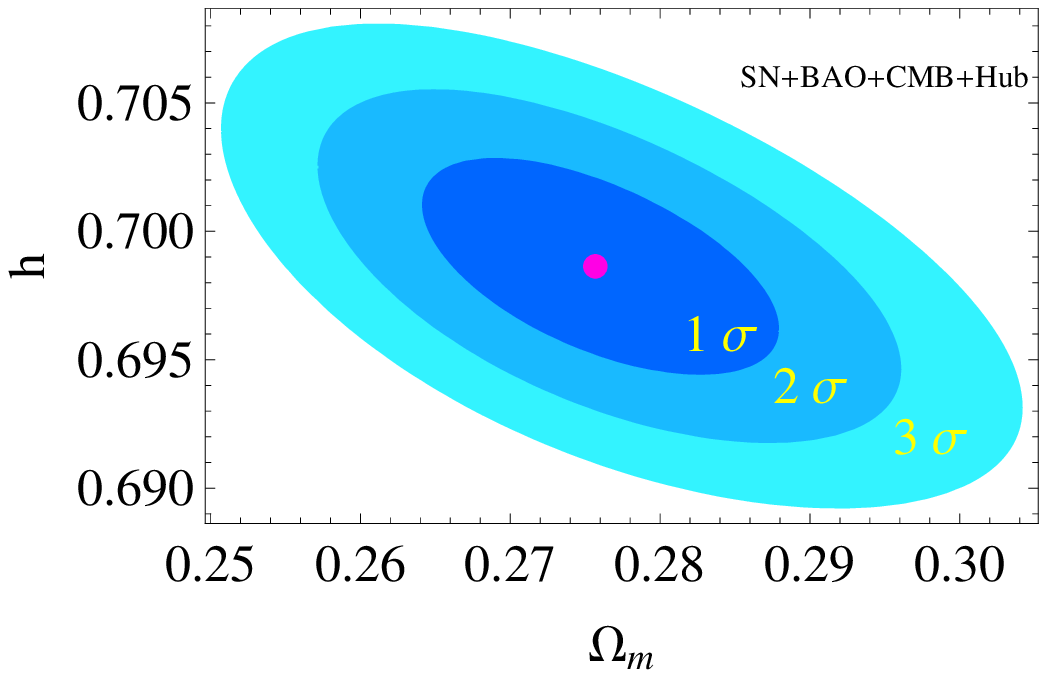}\qquad
\includegraphics[width=6cm]{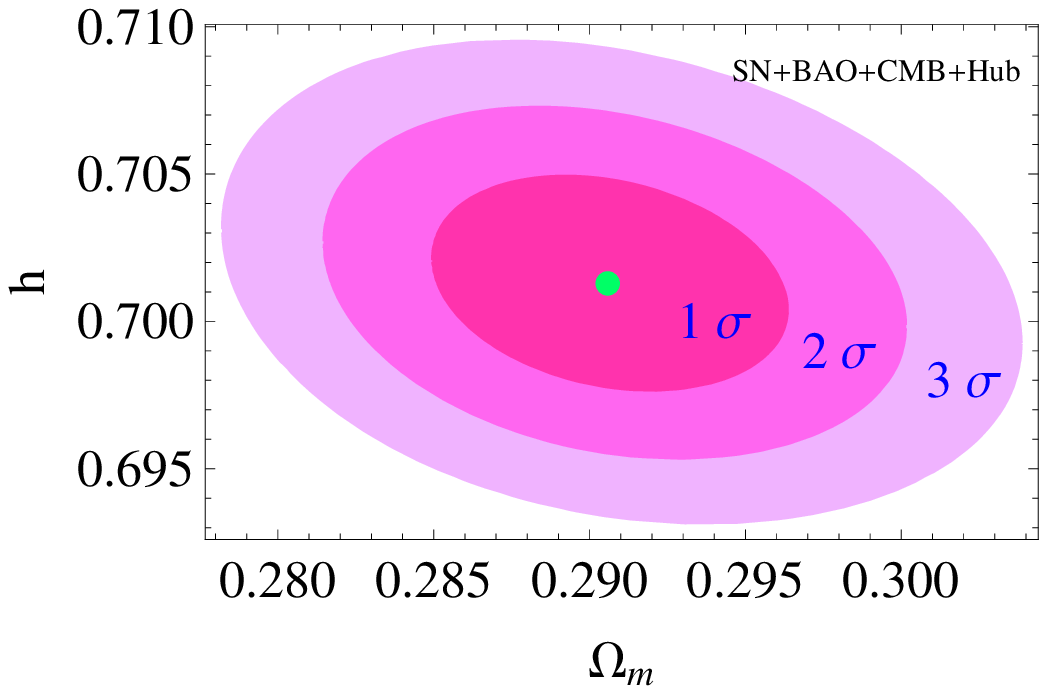}}
  \caption{\label{fig3}
Left: The confidence  contours for the pair
of free parameters ($\Omega_m,h$) obtained by constraining the non-coupled
ESF model with the joint
observational data from
 SN Ia, BAO,   CMB and H(z).
Right:  The confidence contours of the same pair of parameters of  the
 coupled ESF model with $\gamma=0.2$ obtained by the
 same observational data.}
\end{figure}

Thus,  the total $\chi^2$ is  combined   as
\begin{equation}\label{chi2totoal}
\chi^2_{total}=\chi_{SN}^2+\chi_{BAO}^2+\chi_{CMB}^2+\chi_{Hub}^2.
\end{equation}
As the likelihood function is determined as
 $\mathcal{L}\propto {\rm exp}(-\chi_{total}^2/2)$, the best fit values of $\Omega_m$
 and $h$ follow from minimizing Eq.(\ref{chi2totoal}). Fig.\ref{fig3}
 shows the $68.3\%$ ($1\sigma$), $95.4\%$ ($2\sigma$) and $99.7\%$ ($3\sigma$)
 confidence contours  in the $\Omega_m-h$ plane for both
 the ESF model with $\gamma=0$ and $\gamma=0.2$.
For $\gamma=0$, the best fit values of 1-dimension  up to $3\sigma$
confidence level are:
$\Omega_m=0.276\pm0.008^{+0.016+0.024}_{-0.015-0.022}$,
$h=0.699\pm0.003\pm0.006\pm0.008$, with a minimal
$\chi^2_{totoal}=554.713$; while for $\gamma=0.2$, the results are:
$\Omega_m=0.291\pm0.004^{+0.008+0.012}_{-0.007-0.011}$,
$h=0.701\pm0.002\pm0.005\pm0.007$
 with $\chi^2_{totoal}=556.033$.
For comparison,
we also calculate the case of  $\Lambda$CDM and find  it gives almost the same
results  as the non-coupled ESF  model at a very high precision.

Now, we would like to compare these three dark energy models, i.e., $\Lambda$CDM, ESF models with $\gamma=0$ and $\gamma=0.2$. A conventional criterion for comparison
is $\chi^2_{min}/dof$, in which the degree of freedom $dof=N-k$, whereas $N$ and $k$ are the
number of data points and the number of free model parameters, respectively.
We calculated the $\chi^2_{min}/dof$ for the three models, which can be seen in Table. I.
Besides, there are other criterions for model comparison such as the Bayesian evidence \cite{Liddle,Liddle2}.
However, the  Bayesian evidence is usually  sophisticated. As an alternative, we can use
some approximations of Bayesian evidence such as the so-called Bayesian Information Criterion (BIC)
and Akaike Information Criterion (AIC), instead \cite{Weihao}.
The BIC is defined as \cite{Schwarz}
\begin{equation}
{\rm{BIC}}=-2\ln{\mathcal{L}_{max}}+k\ln N,
\end{equation}
and AIC is defined as \cite{Akaike}
\begin{equation}
{\rm{AIC}}=-2\ln{\mathcal{L}_{max}}+2k,
\end{equation}
where $\mathcal{L}_{max}$ is the maximum likelihood.
In the Gaussian cases, $\chi^2_{min}=-2\ln{\mathcal{L}_{max}}$. So, the
differences of BIC and AIC between two models are $\Delta {\rm{BIC}}=\Delta\chi^2
_{min}+\Delta k\ln N$ and $\Delta {\rm{AIC}}=\Delta\chi^2
_{min}+2\Delta k$, respectively.
In Table. I,  we also present the $\Delta {\rm{BIC}}$ and $\Delta {\rm{AIC}}$.
One can find easily from Table. I that, the non-coupled ESF model and $\Lambda$CDM model
 not only
have an almost identical evolution in the recent past ($z<30$),
but also are  undistinguishable
in confronting with the combining observations from SN Ia, BAO, CMB and Hubble parameter.
Moreover, for the coupled ESF model, it only  gives a little
larger $\chi^2_{\rm min}$ and little larger values of all the criterions for model comparison.
Thus, $\gamma>0$ would be favored if further observations support $w<-1$ as indicated in
Refs. \cite{Astier},  since  the coupled and non-coupled ESF models perform similarly  in  $\chi^2$ analysis.

\begin{table}
\caption{\label{table2}
Comparison of the three models considered in this work.
}
\begin{indented}
\lineup
\item[]\begin{tabular}{@{}*{5}{l}}
\br Model & $\chi^2_{total}$ & $\chi^2/dof$ & $\Delta \rm{BIC}$
&$\Delta \rm{AIC}$\cr \mr $\Lambda$CDM  & 554.713 & 0.970 & 0& 0\cr
ESF ($\gamma=0$) &554.713  & 0.970 & 0 &0\cr ESF ($\gamma=0.2$) &
556.033 & 0.972 & 1.32&1.32\cr \br
\end{tabular}
\end{indented}
\end{table}

\section{ Conclusions }

Inspired by a generic feature of effective quantum fields,
we have proposed a scalar field dark energy model,
whose Lagrangian contains a logarithmic correction.
It can be regarded as
a special case of the  generic K-essence models.
For an initial value $\rho_{\phi i}$
ranging over almost infinite orders of magnitude,
$\rho_{\phi }(t)$  tracks radiation and matter,
and the current status ($\Omega_\phi, \Omega_m ) \simeq (0.73, 0.27$)
is always attained.
So the coincidence problem is solved  if
the parameter $\alpha$ is chosen in advance,
but the fine-tuning problem remains.
Moreover, $w$ smoothly crosses $-1$
if the ESF decays into matter.
For a decay rate $\gamma=0.2$,
the EoS arrives at $w_0\simeq -1.05$ at present.
A greater $\gamma$  yields a larger  $\Omega_m$ and
a smaller $w_0$ at present.
As $t \rightarrow \infty$,
the expanding spacetime approaches
the de Sitter as an asymptote,
which is also a stable attractor,
and there is no cosmic big rip.
For the non-coupled  model,
$w$ approaches $-1$ but does not cross $-1$,
and the dynamic behavior
is  almost the same  as $\Lambda$CDM for low redshifts.
In particular, for an initial $\varepsilon_i=0$,
the model reduces to $\Lambda$CDM.
Since the meaning of a non-zero $V(\phi)$  is
unknown, we did not discuss the properties of
the ESF model with $V(\phi)\neq0$. Some particular
forms of $V(\phi)$ would be investigated in the future
study.

In confronting with observations of SN Ia, BAO, CMB and Hubble
parameter, we plotted the confidence contours in the $\Omega_m-h$
plane for the ESF model with $\gamma=0$ and $\gamma=0.2$. The best
fits of the parameters are: $\Omega_m=0.276\pm0.008$ and
$h=0.699\pm0.003$ with $\chi^2_{totoal}=554.713$ for $\gamma=0$;
$\Omega_m=0.291\pm0.004$ and $h=0.701\pm0.002$ with
$\chi^2_{totoal}=556.033$ for $\gamma=0.2$. Furthermore, the
non-coupled ESF model is distinguishable from $\Lambda$CDM model
under present observations. Besides, we compared the three dark
energy models studied in this work using $\chi^2/dof$,
  BIC  and   AIC . It is found that
a non-coupled ESF model is a little more favored, however, the coupled model will survive
if further observations support $w<-1$ strongly.

\section*{Acknowledgments}

 We thank Dr.~Wen Zhao for useful discussions. M.L.
Tong is partially supported by Graduate Student Research Funding
from USTC. Y.Zhang's research work is supported by the CNSF
No.10773009, SRFDP, and CAS.

\end{document}